\documentclass[a4paper,11pt]{article}

\usepackage{bbm}

\usepackage{pos}

\title{A new technique for solving the freezing problem in the complex Langevin simulation of 4D SU(2) gauge theory with a theta term}

\ShortTitle{Complex Langevin simulation of 4D SU(2) gauge theory with a theta term}

\author*[a]{Akira Matsumoto}
\author[b]{Kohta Hatakeyama}
\author[b]{Mitsuaki Hirasawa}
\author[c]{Masazumi Honda}
\author[d]{Yuta Ito}
\author[a,b]{Jun Nishimura}
\author[a]{Atis Yosprakob}

\affiliation[a]{The Graduate University for Advanced Studies, SOKENDAI,\\ 1-1 Oho, Tsukuba, Ibaraki 305-0801, Japan}

\affiliation[b]{KEK Theory Center, High Energy Accelerator Research Organization,\\ 1-1 Oho, Tsukuba, Ibaraki 305-0801, Japan}

\affiliation[c]{Center for Gravitational Physics, Yukawa Institute for Theoretical Physics,\\ Kyoto University, Sakyo-ku, Kyoto 606-8502, Japan}

\affiliation[d]{National Institute of Technology, Tokuyama College,\\ Gakuendai, Shunan, Yamaguchi 745-8585, Japan}

\emailAdd{akiram@post.kek.jp}
\emailAdd{khat@post.kek.jp}
\emailAdd{mitsuaki@post.kek.jp}
\emailAdd{masazumi318@gmail.com}
\emailAdd{y-itou@tokuyama.ac.jp}
\emailAdd{jnishi@post.kek.jp}
\emailAdd{ayosp@post.kek.jp}

\abstract{We apply the complex Langevin method (CLM) to overcome the sign problem
in 4D SU(2) gauge theory with a theta term extending our previous
work on the 2D U(1) case. The topology freezing problem can be solved
by using open boundary conditions in all spatial directions, and the
criterion for justifying the CLM is satisfied even for large $\theta$
as far as the lattice spacing is sufficiently small. However, we find
that the CP symmetry at $\theta=\pi$ remains to be broken explicitly
even in the continuum and infinite-volume limits due to the chosen
boundary conditions. 
In particular, this prevents us from investigating
the interesting phase structures suggested by the 't Hooft anomaly
matching condition. We also try the so-called subvolume method, which
turns out to have a similar problem. We therefore discuss a new technique
within the CLM, which enables us to circumvent the topology freezing
problem without changing the boundary conditions.}

\FullConference{%
 The 38th International Symposium on Lattice Field Theory, LATTICE2021
  26th-30th July, 2021
  Zoom/Gather@Massachusetts Institute of Technology
}

\begin{document}

\begin{flushright}
KEK-TH-2372
\end{flushright}

\maketitle

\section{Introduction}

We can explore the topological nature of quantum field theories via topological terms.
Recently, gauge theories with a theta term have been
studied by \textquoteright t Hooft anomaly matching. 
In particular,
there is a constraint on the phase structure of the 4D SU(N) pure Yang-Mills theory
by a \textquoteright t Hooft anomaly involving the CP and center symmetries at $\theta=\pi$ \cite{Gaiotto:2017yup}.
The constraint is consistent with the well-known scenario at large $N$ \cite{Witten:1980sp},
where the theory at $\theta =\pi$ is confined with spontaneously broken CP at low temperature
and then has a transition to deconfined phase with restored CP at a finite temperature. 
However, it is highly nontrivial 
whether or not this structure persists for small $N$ 
since there are various ways to satisfy the anomaly matching condition.
For instance, the theory for small $N$ at low temperature may be deconfined or gapless
as well as spontaneously broken CP.
Therefore it is an interesting challenge to investigate the phase structure by first-principle calculation at the smallest $N$ i.e.~$N=2$. 
The effect of the theta term is genuinely
non-perturbative. The theory with a theta term should be analyzed by
non-perturbative calculations based on the lattice gauge theory. However,
the Monte Carlo simulation of the theory including the theta term
is difficult due to the sign problem.

The complex Langevin method (CLM) is one of the approaches which allow
us to avoid the sign problem \cite{Klauder:1983sp,Parisi:1984cs,Aarts:2009uq,Aarts:2011ax,Nagata:2015uga,Nagata:2016vkn}.
We use the CLM to study 4D SU(2) gauge theory with the theta term
since its computational cost is cheaper than the other methods.
The topological charge on the 4D lattice is contaminated by short
range fluctuations. Thus, we apply the stout smearing \cite{Morningstar:2003gk}
to recover the topological property. In this method, the effect of
the smearing can be included dynamically. We discuss the behavior
of the topological charge for $\theta\neq0$ in the CLM.

\section{4D SU(2) gauge theory with a theta term}

We consider 4D SU(2) gauge theory on the Euclidean space. The action
for the gauge field $A_{\mu}^{a}$ ($a=1,2,3$) ($\mu=1,\cdots,4$)
is given by
\begin{equation}
S_{g}=\frac{1}{4g^{2}}\int d^{4}xF_{\mu\nu}^{a}F_{\mu\nu}^{a},
\end{equation}
where $g$ is the gauge coupling constant and $F_{\mu\nu}^{a}$ is
the field strength
\begin{equation}
F_{\mu\nu}^{a}=\partial_{\mu}A_{\nu}^{a}-\partial_{\nu}A_{\mu}^{a}-\epsilon^{abc}A_{\mu}^{b}A_{\nu}^{c}.
\end{equation}
The topological charge $Q$ is defined by 
\begin{equation}
Q=\frac{1}{64\pi^{2}}\int d^{4}x\epsilon_{\mu\nu\rho\sigma}F_{\mu\nu}^{a}F_{\rho\sigma}^{a},\label{eq:Q}
\end{equation}
which takes integer values unless the space has a boundary. We introduce
the theta term $S_{\theta}=-i\theta Q$, and thus the action is $S=S_{g}+S_{\theta}$.
This theory has the $2\pi$ periodicity of the parameter $\theta\in\mathbb{R}$,
since the partition function
\begin{equation}
Z=\int\mathcal{D}Ae^{-S_{g}+i\theta Q}
\end{equation}
is invariant under the shift $\theta\rightarrow\theta+2\pi$.

Next, we consider the lattice action for the numerical study. We introduce
link variables $U_{n,\mu}\in\mathrm{SU}(2)$ and define plaquettes.
\begin{equation}
P_{n}^{\mu\nu}=U_{n,\mu}U_{n+\hat{\mu},\nu}U_{n+\hat{\nu},\mu}^{-1}U_{n,\nu}^{-1}
\end{equation}
The index $n$ labels the lattice site and $\hat{\mu}$ represents
the unit vector along the $\mu$-th direction. Note that we use $U_{n,\mu}^{-1}$
instead of $U_{n,\mu}^{\dagger}$ to respect holomorphicity, which
is necessary to justify the CLM.
We define the plaquette action by
\begin{equation}
S_{\beta}=-\frac{\beta}{4}\sum_{n}\sum_{\mu\neq\nu}\mathrm{Tr}P_{n}^{\mu\nu}
\end{equation}
with the coupling constant $\beta$.
For the topological charge on the lattice, we consider the simplest
discretization \cite{DiVecchia:1981aev} given by
the so called \textquotedbl clover leaf\textquotedbl{} formula.
\begin{equation}
Q_{\textrm{cl}}=-\frac{1}{32\pi^{2}}\sum_{n}\frac{1}{2^{4}}\sum_{\mu,\nu,\rho,\sigma=\pm1}^{\pm4}\tilde{\epsilon}_{\mu\nu\rho\sigma}\mathrm{Tr}\left[P_{n}^{\mu\nu}P_{n}^{\rho\sigma}\right]\label{eq:Qcl}
\end{equation}
Here the orientation of the plaquette is generalize to negative directions.
Correspondingly, the anti-symmetric tensor $\tilde{\epsilon}_{\mu\nu\rho\sigma}$
also has negative indices, for example
\begin{equation}
1=\tilde{\epsilon}_{1234}=-\tilde{\epsilon}_{2134}=-\tilde{\epsilon}_{(-1)234}=\cdots.
\end{equation}
Usually the topological charge $Q_{\textrm{cl}}$ does not take integer
values on the lattice due to the discretization effect. We can recover
the topological property of the gauge field by eliminating short-range
fluctuations. Some smoothing techniques, such as the gradient flow,
stout smearing and so on, make the topological charge close to integers.
In this study, we apply the stout smearing to the complex Langevin
method, which is discussed in section \ref{sec:Stout-smearing}.

\section{Complex Langevin method}

Since the theta term is purely imaginary, Monte Carlo studies of the
theory with $\theta\neq0$ is extremely difficult due to the sign
problem. We avoid this problem by using the complex Langevin method
(CLM) \cite{Klauder:1983sp,Parisi:1984cs,Aarts:2009uq,Aarts:2011ax,Nagata:2015uga,Nagata:2016vkn},
which is a generalization of the Langevin method to the system with
a complex action. Its computational cost grows linearly with the system
size, so that we can easily apply the CLM to large systems in a straightforward
manner. In this section, we briefly review how to apply the method
to 4D SU(2) gauge theory.

In the CLM, we consider a fictitious time evolution of the dynamical
variables, which is described by the complex Langevin equation. The
discretized complex Langevin equation for the link variables is given
by
\begin{equation}
U_{n,\mu}(t+\epsilon)=\exp\left[-i\epsilon D_{n,\mu}^{a}S\tau^{a}+i\sqrt{\epsilon}\eta_{n,\mu}(t)\right]U_{n,\mu}(t),
\end{equation}
where $\tau^{a}=\sigma^{a}/2$ are the generators of SU(2). The parameter
$\epsilon\ll1$ is a step size of the discretized fictitious time.
The differential operation $D_{n,\mu}^{a}f$ of the function $f(U)$
with respect to the link variables (Lie group elements) is defined
by
\begin{equation}
D_{n,\mu}^{a}f\left(U_{n,\mu}\right)=\lim_{\epsilon\rightarrow0}\frac{1}{\epsilon}\left[f\left(e^{i\epsilon\tau^{a}}U_{n,\mu}\right)-f\left(U_{n,\mu}\right)\right].
\end{equation}
The term including $D_{n,\mu}^{a}S$ is called the drift term. The
other term is a real Gaussian noise $\eta_{n,\mu}(t)=\eta_{n,\mu}^{a}(t)\tau^{a}$
normalized by
\begin{equation}
\left\langle \eta_{n,\mu}^{a}(t)\eta_{m,\nu}^{b}(t^{\prime})\right\rangle =2\delta_{nm}\delta_{\mu\nu}\delta^{ab}\delta_{tt^{\prime}}.
\end{equation}

The drift term $D_{n,\mu}^{a}S$ is no loner Hermitian for the complex
action. Thus, the link variables deviates from SU(2) in the complex
Langevin simulation. We treat the link variables as SL(2,$\mathbb{C}$)
elements instead of SU(2), which corresponds to complexifying the
gauge field. For the complexified configuration, the drift term and
observables should also be complexified respecting holomorphicity.

The expectation value of $\mathcal{O}$ is calculated from an ensemble
of configurations, which is given by solving the complex Langevin
equation numerically. We can obtain the expectation value $\left\langle \mathcal{O}\right\rangle _{\textrm{CLM}}$
as an average of $\mathcal{O}(U)$ in the ensemble. However, it will
not always agree with the correct expectation value defined by the
path integral. This problem is known as the wrong convergence of the
CLM, which occurs depending on the system, the parameter and the choice
of the dynamical variables. Although we cannot figure out whether the problem
occurs or not a priori, there is a practical criterion for the correct
convergence \cite{Nagata:2016vkn}. We obtain the correct expectation
value $\left\langle \mathcal{O}\right\rangle _{\textrm{CLM}}=\left\langle \mathcal{O}\right\rangle $
only if the probability distribution of the drift term falls off exponentially
or faster. We can easily check the criterion by plotting the histogram
of the magnitude $u$ of the largest drift defined by
\begin{equation}
u=\frac{1}{\sqrt{2}}\max_{n,\mu}\left\Vert D_{n,\mu}^{a}S\tau^{a}\right\Vert,
\label{eq:max_drift}
\end{equation}
where the norm of the matrix is defined by $\left\Vert A\right\Vert ^{2}:=\mathrm{Tr}\left[A^{\dagger}A\right]$.

We can stabilize the complex Langevin simulation by using a technique
called \textquotedbl gauge cooling\textquotedbl{} \cite{Seiler:2012wz}.
The condition of the correct convergence tends to be violated if the
link variables deviates far away from SU(2). The gauge cooling reduces
the non-unitarity of link variables as much as possible. Thus, it
helps the condition to be satisfied. It was also shown that this procedure
does not affect any gauge invariant observable \cite{Nagata:2015uga,Nagata:2016vkn}.
 We apply the gauge cooling at each Langevin step in order to suppress
a rapid growth of non-unitarity.

\section{Stout smearing for the CLM \label{sec:Stout-smearing}}

The theory with a theta term has the $2\pi$ periodicity of $\theta$
, which plays an important role in the appearance of the nontrivial
phase structure at $\theta=\pi$. However, it is difficult to retain
this property on the lattice because the topological charge (\ref{eq:Qcl})
defined by the naive discretization does not takes integer values.
It approaches integers only for the configurations sufficiently close
to the continuum limit. In fact, it is difficult to suppress the short
range fluctuations enough simply by increasing $\beta$.  Thus, we
need a smearing method which makes the configuration sufficiently
smooth even for small $\beta$. In this work, we use the stout smearing
\cite{Morningstar:2003gk}, which is applicable to the CLM. In fact,
its application to the CLM was discussed in the analysis of QCD at
nonzero baryon density \cite{Sexty:2019vqx}. In this section, we
review how to apply the stout smearing to the complex Langevin simulation
of the gauge theory with the theta term.

The procedure of the stout smearing is given by the iteration of the
smearing step, starting from the original configuration $U_{n,\mu}$.
\begin{equation}
U_{n,\mu}=U_{n,\mu}^{(0)}\rightarrow U_{n,\mu}^{(1)}\rightarrow\cdots\rightarrow U_{n,\mu}^{(N_{\rho})}=\tilde{U}_{n,\mu}\label{eq:smearing_steps}
\end{equation}
After $N_{\rho}$ iterations we obtain the smeared configuration $\tilde{U}_{n,\mu}$.
In one (isotropic) smearing step from $k$ to $k+1$, the link variable
$U_{n,\mu}^{(k)}\in\textrm{SL}(2,\mathbb{C})$ is mapped to $U_{n,\mu}^{(k+1)}\in\textrm{SL}(2,\mathbb{C})$
defined by following formulae.
\begin{equation}
U_{n,\mu}^{(k+1)}=e^{iY_{n,\mu}}U_{n,\mu}^{(k)}
\end{equation}
\begin{equation}
iY_{n,\mu}=-\frac{\rho}{2}\mathrm{Tr}\left[J_{n,\mu}\tau^{a}\right]\tau^{a}\label{eq:Y}
\end{equation}
\begin{equation}
J_{n,\mu}=U_{n,\mu}\Omega_{n,\mu}-\bar{\Omega}_{n,\mu}U_{n,\mu}^{-1}
\end{equation}
\begin{equation}
\Omega_{n,\mu}=\sum_{\sigma(\neq\mu)}\left(U_{n+\hat{\mu},\sigma}U_{n+\hat{\sigma},\mu}^{-1}U_{n,\sigma}^{-1}+U_{n+\hat{\mu}-\hat{\sigma},\sigma}^{-1}U_{n-\hat{\sigma},\mu}^{-1}U_{n-\hat{\sigma},\sigma}\right)\label{eq:OMG}
\end{equation}
\begin{equation}
\bar{\Omega}_{n,\mu}=\sum_{\sigma(\neq\mu)}\left(U_{n,\sigma}U_{n+\hat{\sigma},\mu}U_{n+\hat{\mu},\sigma}^{-1}+U_{n-\hat{\sigma},\sigma}^{-1}U_{n-\hat{\sigma},\mu}U_{n+\hat{\mu}-\hat{\sigma},\sigma}\right)\label{eq:OMGbar}
\end{equation}
The parameter $\rho>0$ should be chosen appropriately, depending
on the system.

We use the topological charge (\ref{eq:Qcl}) calculated from the
smeared configuration $\tilde{U}_{n,\mu}$ 
\begin{equation}
Q:=Q_{\textrm{cl}}(\tilde{U})\label{eq:def_Q}
\end{equation}
to define the theta term $S_{\theta}=-i\theta Q$ on the lattice.
For the complex Langevin simulation, we need to calculate the drift
term $D_{n,\mu}^{a}S_{\theta}$ from the theta term. Although $S_{\theta}$
is a complicated function of the original link variable $U_{n,\mu}$,
it is possible to calculate the drift force 
\begin{equation}
F_{n,\mu}=i\tau^{a}D_{n,\mu}^{a}S_{\theta}\label{eq:drift_theta}
\end{equation}
by reversing the smearing steps (\ref{eq:smearing_steps}). We define
the drift force for the link variables $U_{n,\mu}^{(k)}$ as
\begin{equation}
F_{n,\mu}^{(k)}=i\tau^{a}D_{n,\mu}^{(k)a}S_{\theta},
\end{equation}
where $D_{n,\mu}^{(k)a}$ represents a differential operation with
respect to $U_{n,\mu}^{(k)}$. As a first step to calculate (\ref{eq:drift_theta}),
the calculation of the drift force $\tilde{F}_{n,\mu}=F_{n,\mu}^{(N_{\rho})}$
for the smeared link $\tilde{U}_{n,\mu}=U_{n,\mu}^{(N_{\rho})}$ is
straightforward. Once we obtain the initial drift force $\tilde{F}_{n,\mu}$,
the subsequent ones are given by the map from $F_{n,\mu}^{(k)}$ to
$F_{n,\mu}^{(k-1)}$ iteratively.
\begin{equation}
\tilde{F}_{n,\mu}=F_{n,\mu}^{(N_{\rho})}\rightarrow F_{n,\mu}^{(N_{\rho}-1)}\rightarrow\cdots\rightarrow F_{n,\mu}^{(0)}=F_{n,\mu}
\end{equation}
The map of the drift force is given by the following formulae,
where the final step from $F_{n,\mu}^{\prime}=F_{n,\mu}^{(1)}$
to $F_{n,\mu}=F_{n,\mu}^{(0)}$ is shown as an example.
\begin{equation}
F_{n,\mu}=e^{-iY_{n,\mu}}F_{n,\mu}^{\prime}e^{iY_{n,\mu}}+\rho\mathrm{Tr}\left[(U_{n,\mu}M_{n,\mu}+\bar{M}_{n,\mu}U_{n,\mu}^{-1})\tau^{a}\right]\tau^{a}
\end{equation}
\begin{align}
M_{n,\mu} & =-\Omega_{n,\mu}\Lambda_{n,\mu}\nonumber \\
 & +\sum_{\nu(\neq\mu)}\left[U_{n+\hat{\mu},\nu}U_{n+\hat{\nu},\mu}^{-1}(U_{n,\nu}^{-1}\Lambda_{n,\nu}+\Lambda_{n+\hat{\nu},\mu}U_{n,\nu}^{-1})\right.\nonumber \\
 & \hphantom{=\sum_{\nu(\neq\mu)}}+U_{n+\hat{\mu}-\hat{\nu},\nu}^{-1}U_{n-\hat{\nu},\mu}^{-1}(\Lambda_{n-\hat{\nu},\mu}-\Lambda_{n-\hat{\nu},\nu})U_{n-\hat{\nu},\nu}\nonumber \\
 & \hphantom{=\sum_{\nu(\neq\mu)}}\left.-\Lambda_{n+\hat{\mu},\nu}U_{n+\hat{\mu},\nu}U_{n+\hat{\nu},\mu}^{-1}U_{n,\nu}^{-1}+U_{n+\hat{\mu}-\hat{\nu},\nu}^{-1}\Lambda_{n+\hat{\mu}-\hat{\nu},\nu}U_{n-\hat{\nu},\mu}^{-1}U_{n-\hat{\nu},\nu}\right]
\end{align}
\begin{align}
\bar{M}_{n,\mu} & =-\Lambda_{n,\mu}\bar{\Omega}_{n,\mu}\nonumber \\
 & +\sum_{\nu(\neq\mu)}\left[(\Lambda_{n,\nu}U_{n,\nu}+U_{n,\nu}\Lambda_{n+\hat{\nu},\mu})U_{n+\hat{\nu},\mu}U_{n+\hat{\mu},\nu}^{-1}\right.\nonumber \\
 & \hphantom{=\sum_{\nu(\neq\mu)}}+U_{n-\hat{\nu},\nu}^{-1}(\Lambda_{n-\hat{\nu},\mu}-\Lambda_{n-\hat{\nu},\nu})U_{n-\hat{\nu},\mu}U_{n+\hat{\mu}-\hat{\nu},\nu}\nonumber \\
 & \hphantom{=\sum_{\nu(\neq\mu)}}\left.-U_{n,\nu}U_{n+\hat{\nu},\mu}U_{n+\hat{\mu},\nu}^{-1}\Lambda_{n+\hat{\mu},\nu}+U_{n-\hat{\nu},\nu}^{-1}U_{n-\hat{\nu},\mu}\Lambda_{n+\hat{\mu}-\hat{\nu},\nu}U_{n+\hat{\mu}-\hat{\nu},\nu}\right]
\end{align}
\begin{equation}
\Lambda_{m,\nu}=\mathrm{Tr}\left[\hat{\Lambda}_{m,\nu}\tau^{b}\right]\tau^{b}
\end{equation}
\begin{equation}
\hat{\Lambda}_{m,\nu}=-\frac{1}{2\kappa_{m,\nu}^{2}}\left(1-\frac{\sin2\kappa_{m,\nu}}{2\kappa_{m,\nu}}\right)\mathrm{Tr}\left[F_{m,\nu}^{\prime}iY_{m,\nu}\right]iY_{m,\nu}+\frac{\sin\kappa_{m,\nu}}{\kappa_{m,\nu}}e^{-iY_{m,\nu}}F_{m,\nu}^{\prime}
\end{equation}
\begin{equation}
\kappa_{n,\mu}=\sqrt{-\det Y_{n,\mu}}
\end{equation}
Note that $Y_{n,\mu}$, $\Omega_{n,\mu}$ and $\bar{\Omega}_{n,\mu}$
are defined by (\ref{eq:Y}), (\ref{eq:OMG}) and (\ref{eq:OMGbar})
respectively. They are calculated from $U_{n,\mu}$ in this case.
The drift term calculated in this way respects the holomorphicity.
The calculation time and the memory size required for the simulation
are proportional to the number of steps $N_{\rho}$.

\section{Result of the CLM \label{sec:Result}}

In this section, we show the results of the complex Langevin simulation.
So far, we have found that the CLM using the naive definition (\ref{eq:Qcl})
of the topological charge without the smearing works in the high-temperature
region (deconfined phase). As a first step, we focus on the high-temperature
region and try to see the effect of the stout smearing on the topological
charge. 

\begin{figure}
\begin{centering}
\includegraphics[scale=0.6]{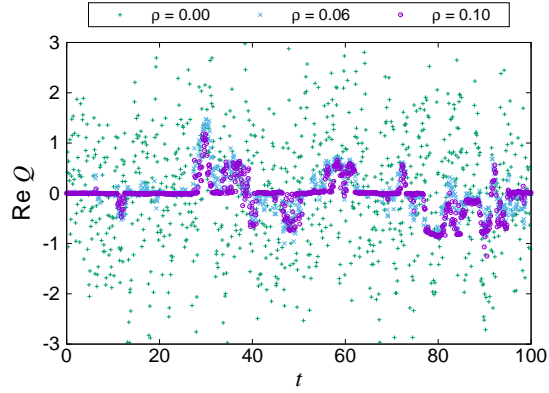} 
\par\end{centering}
\caption{\label{fig:history_Q} The history of the topological charge defined
by (\ref{eq:def_Q}) in the Langevin simulation for $\theta=0$. The
lattice size is $24^{3}\times4$, and the coupling constant is $\beta=2.5$.
The horizontal axis is the fictitious time $t$ of the Langevin simulation.}
\end{figure}

Before introducing the theta term, we check the effect of the smearing
by changing the smearing parameters for $\theta=0$. The number of
steps $N_{\rho}$ and the step size $\rho$ should be large enough
to eliminate the short range fluctuations. However, it is difficult
to increase $N_{\rho}$ a lot since the calculation time and the memory
size increase with $N_{\rho}$. If $\rho$ is too large, the nontrivial
topological excitation will be destroyed. For $\beta>2.4$, which
corresponds to the high-temperature region in our setup, we find that
$N_{\rho}=20$ is enough to recover the topological property. In figure
\ref{fig:history_Q}, we show the history of the topological charge
defined by (\ref{eq:def_Q}) in the real Langevin simulation for $\theta=0$.
There are three series of data with $\rho=0$, 0.06 and 0.1. We plot
the topological charge without the smearing namely $\rho=0$ for comparison.
The topological charge with $\rho=0$ is noisy, and it is difficult
to see the topological property. Once we introduce the smearing, we
can see the transitions between the topological sectors clearly.

\begin{figure}
\begin{centering}
\includegraphics[scale=0.6]{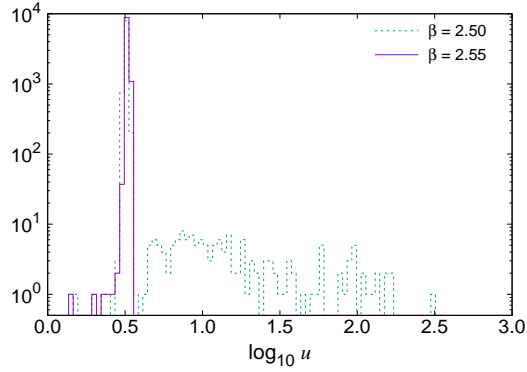} 
\par\end{centering}
\caption{\label{fig:drift_histogram} The histogram of the maximum drift term
(\ref{eq:max_drift}) for $\theta=\pi/4$ in log scale. The horizontal
axis is $\log_{10}u$. The lattice size is $24^{3}\times4$, and the
smearing parameters are $N_{\rho}=20$ and $\rho=0.06$.}
\end{figure}

Next, we show the results of the complex Langevin simulation for $\theta=\pi/4$.
In this simulation, the lattice size is $24^{3}\times4$, and the
smearing parameters are $N_{\rho}=20$ and $\rho=0.06$. In figure
\ref{fig:drift_histogram}, we show the histogram of the magnitude
$u$ of the largest drift term defined in (\ref{eq:max_drift}). The
distribution falls off rapidly for $\beta=2.55$, but it does not
for $\beta=2.5$. Thus, the criterion for correct convergence is satisfied
only for $\beta=2.55$. Typically, the coupling constant $\beta$
should be large enough to satisfy the criterion. We found that the
CLM works if $\beta\gtrsim2.55$ for $\theta=\pi/4$ on the $24^{3}\times4$
lattice.

\begin{figure}

\centering{}\includegraphics[scale=0.6]{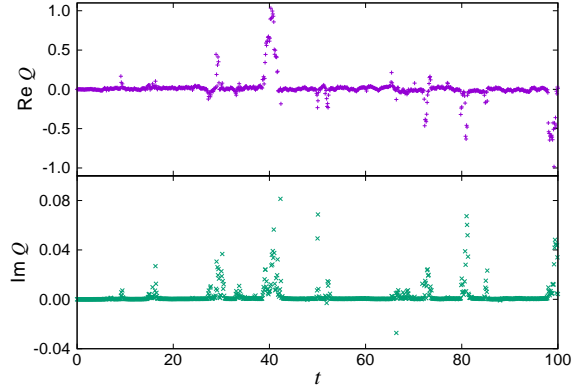}
\caption{\label{fig:history_complexQ} The history of the topological charge
for $\theta=\pi/4$. The upper plot show the real part and the lower
plot show the imaginary part. The lattice size is $24^{3}\times4$,
and the coupling constant is $\beta=2.55$. The horizontal axis is
the fictitious time $t$ of the Langevin simulation.}
\end{figure}

In figure \ref{fig:history_complexQ}, we show the history of the
topological charge for $\beta=2.55$. Since the gauge group is extended
to SL(2,$\mathbb{C}$) in the CLM, the topological charge has an imaginary
part in general. We plot both of the real part and the imaginary part.
There are some topological excitations in the history of $\mathrm{Re}Q$.
The imaginary part vanishes after the smearing in most cases, but
it increases rapidly when the real part changes.

The expectation value of the topological charge has a nonzero imaginary
part if CP is broken. Since the theta term breaks CP explicitly for
$\theta/\pi\notin\mathbb{Z}$, it is consistent that $\mathrm{Im}Q$
becomes nonzero in our simulation. We find that the fluctuation of
$\mathrm{Re}Q$ is necessary to obtain the nonzero $\mathrm{Im}Q$.
Indeed, the imaginary part are close to zero while the configuration
stays in a single topological sector.

We also find that the rapid growth of $\mathrm{Im}Q$ makes the simulation
unstable. The imaginary part originates from the non-unitarity of
the configuration, which can be a source of the large drift. We need
to set $\beta$ large enough to avoid this problem. However, the fluctuation
of $Q$ is highly suppressed for larger $\beta$, and the autocorrelation
time of $Q$ becomes longer than the simulation time. It is known
as freezing of the topological charge, which causes a problem with
the ergodicity. Therefore, it is difficult to avoid the large drift
simply by increasing $\beta$ further.

\section{Summary}

The sign problem prevents us from studying gauge theories with a theta
term by the Monte Carlo simulation. In this work, we applied the complex
Langevin method (CLM) to 4D SU(2) gauge theory to avoid the problem.
We found that the criterion for correct convergence of the CLM is
satisfied in the high temperature region. However, the naively defined
topological charge does not take integer values due to the contamination
by short range fluctuations.
For this reason, we introduce the stout smearing
in the CLM in order to recover the topological property. The effect
of the smearing can be included in the Langevin dynamics itself as
well as in observables. We confirmed that the real part of the topological
charge becomes close to an integer after the smearing. On the other
hand, the imaginary part vanishes mostly, but it grows rapidly as
the real part changes. This behavior is consistent with the topological
nature of the theory, although it is difficult to deal with in the
numerical simulation.

We need to increase $\beta$ to suppress the large drift. On the other
hand, we cannot increase it due to the topology freezing. It seems
to be necessary to resolve either of the topology freezing or the
large drift in the CLM. However, it is possible that the appearance
of large drift is related to the topology change, as we found in our
previous study of 2D U(1) gauge theory \cite{Hirasawa:2020bnl}. In
that case, we need to modify the boundary condition
or try some possible ways to suppress the large drifts,
such as improving the gauge cooling or the smearing method.

\acknowledgments
We would like to thank
R.~Kitano, N.~Yamada, T.~Ishikawa and Y.~Tanizaki for valuable discussions.
The computations were carried out on the PC clusters in KEK Computing Research Center and KEK Theory Center.
This work is supported by the Particle, Nuclear and Astro Physics Simulation Program No.2020-009 (FY2020) and No.2021-005 (FY2021) 
of Institute of Particle and Nuclear Studies, High Energy Accelerator Research Organization (KEK).
The work of M.~Honda is supported by MEXT Q-LEAP, JST PRESTO Grant Number JPMJPR2117
and JSPS Grant-in-Aid for Transformative Research Areas (A) JP21H05190.

\bibliographystyle{JHEP}
\bibliography{ref}

\end{document}